\documentclass{ar2e}%%%Put [range] option here for references citations to
\usepackage{amsmath}
\usepackage{amssymb}
\usepackage{graphicx}
\usepackage[usenames]{color}
\usepackage{collref}

\begin{document}

\newcommand{\ahat}{(a/M)}
\newcommand{\etal}{et al. \,}
\newcommand{\kms}{${\rm km s}^{-1}$}
\newcommand{\MSun}{M_{\odot}}

\newcommand{\jb}[1]{{ \textcolor{magenta}{ ({\em  {#1} -- JB})} } }
\newcommand{\bjk}[1]{{ \textcolor{JungleGreen}{ ({\em  {#1} -- BJK})} } }

\jname{Annu. Rev. Nucl. Part. Sci.}
\jyear{XXXX}
\jvol{YY}
\ARinfo{1056-8700/97/0610-00}

\title{The Final Merger of Black-Hole Binaries}

\markboth{Final Merger of Black-Hole Binaries}{Final Merger of Black-Hole Binaries}

\author{Joan Centrella$^1$, John G. Baker$^2$, Bernard J. Kelly$^3$, and James R. van Meter$^4$ \affiliation{$^1$Gravitational Astrophysics Laboratory, NASA GSFC, 8800 Greenbelt Rd., Greenbelt, MD 20771, USA\\email: Joan.M.Centrella@nasa.gov; corresponding author\\$^2$Gravitational Astrophysics Laboratory, NASA GSFC, 8800 Greenbelt Rd., Greenbelt, MD 20771, USA\\email: John.G.Baker@nasa.gov\\$^3$CRESST \& Gravitational Astrophysics Laboratory, NASA/GSFC, 8800 Greenbelt Rd., Greenbelt, MD 20771, USA\\Dept. of Physics, University of Maryland, Baltimore County, 1000 Hilltop Circle, Baltimore, MD 21250, USA\\email: Bernard.J.Kelly@nasa.gov\\$^4$CRESST \& Gravitational Astrophysics Laboratory, NASA/GSFC, 8800 Greenbelt Rd., Greenbelt, MD 20771, USA\\Dept. of Physics, University of Maryland, Baltimore County, 1000 Hilltop Circle, Baltimore, MD 21250, USA\\email: James.R.VanMeter@nasa.gov\\}}


\begin{abstract}
Recent breakthroughs in the field of numerical relativity have led to 
dramatic progress in understanding the predictions of General Relativity
for the dynamical interactions of two black holes in the regime of
very strong gravitational fields.  
Such black-hole binaries are important astrophysical systems and are a
key target of current and developing gravitational-wave detectors.
The waveform signature of strong gravitational radiation emitted 
as the black holes fall together and merge provides a clear observable record
of the process.  
After decades of slow progress, these mergers and the
gravitational-wave signals they generate can now be routinely calculated
using the methods of numerical relativity.
We review recent advances in understanding the predicted physics of
events and the consequent radiation, and discuss some of the
impacts this new knowledge is having in various areas of astrophysics. 
\end{abstract}

\maketitle

\section{Introduction}
\label{sec:intro}

\subsubsection{Black Holes}
\label{sssec:intro_bhs}

It has been nearly a century since Albert Einstein's profound
physical insight revealed our current standard model of gravitational
physics, General Relativity.  Among the theory's 
extraordinary consequences was the predicted existence 
of black holes, nonlinearly self-gravitating, stable objects in which
gravitational forces have completely overcome all other physical
interactions.  Though once considered a mathematical curiosity,
General relativity's 
description of black holes now provides the best explanation
for a widespread class of astrophysical objects whose gravitational
potential wells power many of the most energetic astronomical phenomena
observed.  These range from stellar black-holes powering 
X-ray sources in the neighboring regions of our galaxy, to supermassive
monsters with masses $\sim (10^6 - 10^9)\MSun$, 
where $\MSun$ is the mass of the Sun,
with far-reaching astrophysical consequences.

\subsubsection{Gravitational Wave Observations}
\label{sssec:intro_gws}

In the coming decade, anticipated observations of gravitational waves
from black hole binaries will open a new window
onto the universe.  Interpreting such observations will
richly engage aspects of our theoretical 
understanding of strong-field gravity which have never before
been confronted with empirical observations. 

The final coalescence of binaries consisting of two comparable-mass
black-holes, with mass ratios $q = M_1/M_2 \sim (1-10)$, is
expected to be one of the strongest astrophysical sources of energy 
 the form of gravitational radiation. As the emission of gravitational waves
removes energy and momentum from the binary, the orbits shrink and the black holes
eventually merge together into a single black hole, producing an
intense burst of radiation.  With the advent of ground-based
gravitational wave detectors such as LIGO and VIRGO (which will detect
mergers of black holes with masses in the range 
$\sim (10 - 100)\MSun$), and with
planning underway for the space-based LISA (which will observe mergers
of massive black holes, with masses 
$\sim (10^4 - 10^6)\MSun$), knowledge of 
black-hole binary gravitational waveforms is urgent.

\subsubsection{Black-Hole Binary Coalescence}
\label{sssec:intro_bhbmergers}

Black-hole binary coalescence
can be thought of as proceeding in three stages.
During the inspiral
the holes have wide enough separations that they can be treated as point particles.
In this stage, the orbital period 
 is much shorter than the timescale on 
which the orbital parameters change, and the holes spiral together on quasi-circular orbits.
When the holes get so close together that they can no longer be approximated as 
point particles, they enter the merger phase. 
In this strong-field, dynamical regime,
the black holes plunge together and merge into a single, 
highly distorted black hole, surrounded by a common horizon.
This remnant black hole then ``rings down,'' shedding
its nonaxisymmetric modes through 
gravitational wave emission and settling down into a quiescent rotating black hole. 

\subsubsection{Gravitational Waves from the Coalescence}
\label{sssec:intro_mergerwfs}

Knowledge of the gravitational waveforms from these three stages
of black hole coalescence is important for gravitational wave
detection and data analysis, as well as astrophysical applications.
The inspiral can be calculated analytically 
using the post-Newtonian (PN) approximation,
which is an expansion of the full equations of General Relativity
in powers of $\epsilon \sim v^2/c^2 \sim G M/(Rc^2)$, where $v$ is the
characteristic velocity of the source, 
$M$ is its mass, and $R$ is its characteristic size.
The inspiral gravitational waveform is a ``chirp'', which is a sinusoid 
increasing in both frequency and amplitude.
The ringdown can also be calculated analytically using techniques of 
black-hole perturbation theory, and 
the resulting gravitational waveforms are damped sinusoids.
However, the merger stage can only be understood using full numerical simulations of 
the Einstein equations in 3-D,
and the resulting gravitational waveforms were unknown -- until recently.

The final merger of comparable-mass black holes is a powerful source
of gravitational radiation. The
gravitational waveforms emitted by a black-hole
binary with mass ratio $q$ scale with the total mass $M$ of the
binary. Setting $c=1$ 
and $G=1$, 
we can express both length and time scales in terms of the total
mass: $1 M \sim 5 \times 10^{-6} (M/\MSun) {\rm sec} 
\sim 1.5 (M/\MSun) {\rm km}$. 
The strong-field merger will produce a burst of
gravitational radiation 
lasting $\sim 100M$ and having a luminosity $\sim 10^{23}L_{\odot}$,
which is greater than the combined luminosities of all the stars
in the visible universe. For stellar black-hole binaries ($M \sim 10\MSun$) 
this luminosity will last for $\sim 5$ ms, 
and for massive black-hole binaries 
($M \sim 10^6 \MSun$), for $\sim 10$ min.

\subsubsection{Calculating the Merger}
\label{sssec:intro_mergercalc}

Understanding the final merger of a black-hole binary
requires solving the full Einstein equations 
using the methods of numerical relativity.   
This endeavor has proved to have many challenging aspects,
ranging from providing astrophysically relevant initial data to
understanding the structure of the Einstein equations, and 
the solution eluded researchers for many years. Recently, 
a series of dramatic breakthroughs has ignited the field, making 
robust, stable, and accurate simulations of binary mergers possible
for the first time. These models are opening our understanding of
strong-field dynamics, impacting gravitational wave searches 
and other areas of astrophysics.  In this article we 
provide an overview of
these exciting developments, highlighting the key physical results
that are emerging.

\section{Steps toward a robust black-hole binary model}
\label{sec:bhbmodel}

Computing the strong-field merger of two comparable-mass black
holes has a long history, with the first attempt dating back
more than 40 years.  Overall, progress was generally slow and
incremental, requiring the interplay among new ideas in black hole
modeling, mathematical investigations into the structure of
the Einstein equations, the development of effective gauge conditions,
advances in computational techniques, and the availability of
high performance computing resources.  Here we provide a brief
review of key developments in this arena.

\subsection{Overview of Numerical Relativity Issues}
\label{ssec:bhbmodel_nr}

Solving Einstein's equations 
on a computer typically requires slicing 4-D spacetime into a stack 
of 3-D spacelike hypersurfaces, each labeled by time $t$ 
\cite{Arnowitt:1962hi,Misner73,Alcubierre08}. The Einstein equations then 
divide into two sets: constraint and evolution
equations. The constraints are a set of time-independent elliptical equations
that must hold on each slice. In particular, the constraints are solved first  
to obtain valid initial data for a black-hole binary simulation. 
This data is then 
propagated forward in time using the evolution 
equations.  

\subsubsection{Gauge Freedom}
\label{sssec:bhbmodel_nr_gauge}

General Relativity has four spacetime coordinate
degrees of freedom, which are associated with four freely-specifiable 
coordinate or gauge functions that govern
the future development of 
both time and the spatial coordinates. These are generally taken to be
the lapse function $\alpha$, which gives the lapse of proper
time $(\alpha\,\delta t)$ between neighboring slices, and the shift vector
$\beta^i$, which determines how the
spatial coordinates develop from one slice to the next.
Appropriate choices for the lapse and shift have proven
to be crucial for successful black hole evolutions
\cite{Smarr:1977uf,Alcubierre:2002kk}. 

\subsubsection{Formulation of the Evolution Equations}
\label{sssec:bhbmodel_nr_formulation}

For many years, a primary challenge of numerical relativity was simply
to decide which equations to solve \cite{Alcubierre08}. 
There is no unique formulation of the Einstein equations; rather
there are many choices regarding, for example, which variables to use,
whether to write the equations as first-order or second-order in time,
and which coordinate conditions to impose.
These choices are not arbitrary because some formulations turn out to be more 
``numerically friendly'' than others.
That is, although they are analytically equivalent when constraints and
auxiliary variable definitions are assumed exact, some formulations may be 
unstable in practice.
For example,the evolution equations may admit rapidly growing solutions 
which violate the constraints. In this case, although the initial data 
may contain only tiny errors, the subsequent evolution
may produce violations of Einstein's equations which ``blow-up''. 
There can also be pathological numerical solutions that don't represent
solutions to the analytic evolution equations at all, but are supported
by the discrete numerical grid structure and depend on the details of
the finite-differencing operators.  

\subsubsection{Numerical Methods}
\label{ssec:bhbmodel_nr_methods}

Once a formulation has been chosen, the Einstein equations are
solved numerically for various field variables on a grid of
discrete points that represents the spacetime domain of interest.
There are two general approaches
for dealing with the spatial derivatives that appear in these equations.  
Finite differencing interpolates the derivative at a
given point from the surrounding points according to a Taylor expansion in
the grid spacing.  Spectral methods assume a solution in the form of a
summation of orthogonal functions; once the coefficients are obtained
numerically, the derivatives can be found analytically. 
 In both cases, 
time integration can be handled in a number of ways, most commonly by a 
Runge-Kutta algorithm.
To date, most numerical relativity solutions have been carried out
using finite differencing, with most current work employing 
a Cartesian grid in three spatial dimensions, 
although results from evolutions using
spectral methods are now becoming available.

\subsubsection{initial conditions for black holes}
\label{ssec:bhbmodel_nr_id}

To model astrophysical binaries, initial data must be generated 
for two black holes moving on 
quasicircular inspiralling orbits and having
a mass-ratio $q$ and some configuration of
spins. One usually
conceptualizes the initial state of the system with a particle-like
description of the masses, positions, momenta, and spins of each black
hole.  Since general relativity is a field theory, such a description
can only be seen as a first step.  The model requires a full
description for the initial field configuration which satisfies
General Relativity's four initial value constraints, and which somehow
corresponds to the system we have described in these particle-like
terms.

\subsubsection{Evolving the Binary}
\label{sssec:bhbmodel_nr_mesh}

This system is evolved for several orbits and then
through plunge, merger, and ringdown, for a total duration
on the order of several hundred $M$ or more.
To obtain the gravitational waveforms, the
radiation must be extracted
 from the simulation at large enough distances
from the source to be in the ``wave zone''. Since the
 length scales of the black holes are $\sim M$
and the wavelengths of the emitted radiation
are $ \sim (10-100)M$, it is clear that some sort of
variable resolution such as adaptive mesh refinement
must be used to handle the very large computational domains needed.

\subsection{The Lazarus Approach: Hybrid Black Hole Merger Waveforms}
\label{ssec:bhbmodel_lazarus}

By the late 1990's numerical relativity had developed to the point
that brief  simulations of black-hole binaries in three spatial dimensions
plus time were possible.  These techniques
were sufficient for evolving 
promptly merging ``grazing collisions'' 
of black holes~\cite{Bruegmann:1997uc,Alcubierre:2000ke}.  However, the
simulations were not indefinitely stable, but rather would typically crash
after $\sim (10-30)M$, well before any significant portion of a binary
orbit could be evolved.  

\subsubsection{Hybrid simulations}
\label{ssec:bhbmodel_lazarus_hybrid}

In this arena, the Lazarus method emerged as a novel approach to obtaining
black-hole binary waveforms, combining short numerical relativity 
simulations with black hole perturbation methods
\cite{Baker:2000zm,Baker:2001sf}.
Since numerical relativity codes were then only
able to evolve for a brief period, and since perturbation theory could
approximate the late time dynamics of the distorted remnant black 
hole, the Lazarus Project aimed to apply numerical relativity to evolve
the strong-field approach to merger,  thereby
providing a hybrid model for a significant part of the problem.
Starting from quasi-circular initial configurations near the start of
the final merger,
the black holes were evolved using numerical relativity for $\sim 15M$
until just before the simulation became inaccurate.  Then, via a
complicated interface, data from late in the numerical simulations was
interpreted as initial data describing the perturbed final black hole
and the emerging radiation.  Finally, black hole perturbation theory techniques
were applied to evolve this data and derive the full waveforms.

\subsubsection{First Results}
\label{sssec:bhbmodel_lazarus_results}

\begin{figure}
\includegraphics*[width=5.0in]{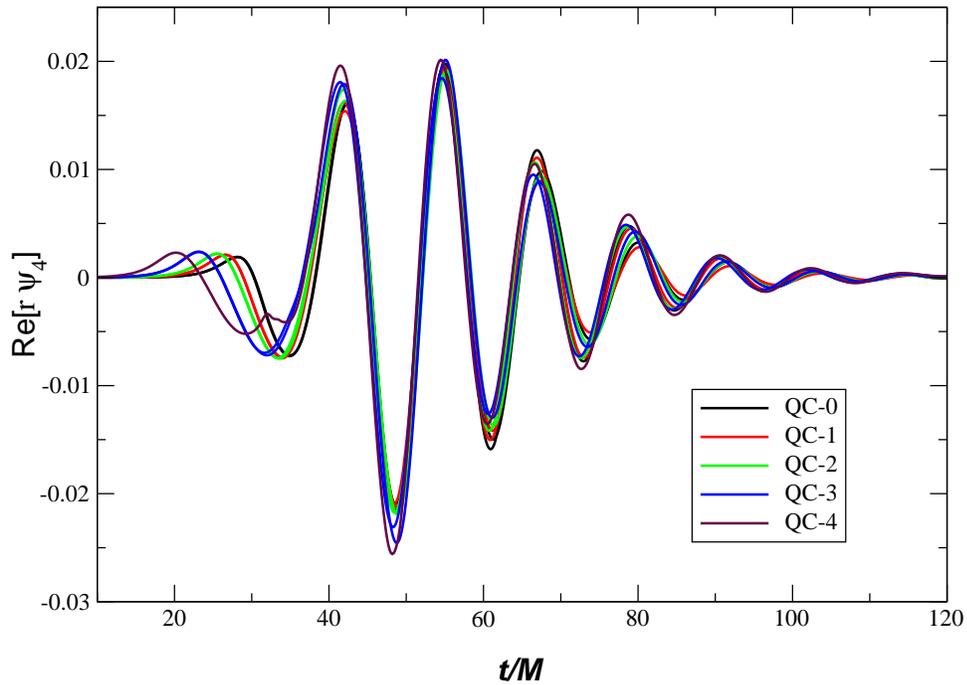}
\caption{Waveforms from Lazarus models of
 equal-mass, nonspinning black hole
mergers. This figure shows the
$\ell=2, m=2$ component of
Re($r \Psi_4$), corresponding to one of the two polarization
states of the emitted gravitational radiation.  Curves are plotted for 10
simulations having different initial black hole separations (designated
QC-0, etc.) and transition times to perturbative evolution.
Reprinted
from \cite{Baker:2002qf} and copyright 2002 by the American Physical
Society
(http://link.aps.org/abstract/PRD/v65/e124012).
}
\label{Fig1_LazarusWF}
\end{figure}

The Lazarus simulations gave the first indication of what
might be expected for the terminal burst of radiation from coalescing
black-hole binaries. Figure~\ref{Fig1_LazarusWF} shows that 
the dominant $\ell=2,m=2$ spin-(-2)-weighted
spherical harmonic component has a brief burst of radiation smoothly
joining the damped sinusoidal signal of the ringdown.
The waveforms were
remarkable for their simplicity, with predominantly circular polarization
(in the $\ell = 2, m=2$ mode), and a steady evolution of polarization frequency
and amplitude.  Since these characteristics were robust
under variations in the model, it was conjectured that the 
waveforms from a binary starting from a wide
separation late in the inspiral 
would be well characterized by these simple features.  
Subsequently,
the Lazarus method was applied to study mergers of more
generic black-hole binaries \cite{Baker:2003ds,Campanelli:2004zw}.

\subsection{Toward Evolving a Black-Hole Binary Orbit}
\label{ssec:bhbmodel_evol}

\subsubsection{The BSSN system}
\label{sssec:bhbmodel_evol_bssn}

Concurrently with the Lazarus investigations, further progress was
being made toward full numerical relativity simulations of black-hole
binary mergers.  One major milestone was the development of a conformal
formulation of the Einstein equations known as the 
Baumgarte-Shapiro-Shibata-Nakamura or ``BSSN'' system
\cite{Shibata:1995we,Baumgarte:1998te}, that overcame some instability
problems associated with an earlier formulation in use at the time
\cite{Arnowitt:1962hi,York79}.
In this approach,
the set of evolution equations is written with first-order 
time derivatives and second-order spatial
derivatives, and is strongly hyperbolic \cite{Nagy:2004td,Reula:2004xd}.
Stable time evolution was accomplished using a coordinate condition
that evolves the lapse function dynamically, causing the slices to avoid
crashing into the black hole singularities
\cite{Anninos:1995am,Alcubierre:2000ke}.
However the simulations were still limited to durations
$\lesssim (30-40)M$ of stable evolution, by a failure
of the spatial coordinate system known as ``grid stretching,''
in which the coordinates tend to fall into the black holes,
and by instabilities related to how the black holes were handled
numerically.

\subsubsection{Controlling the spatial coordinates} 
\label{sssec:bhbmodel_evol_spatial}

Eliminating grid-stretching required developing appropriate techniques
for dynamically controlling the spatial coordinates, which are governed
by the evolution of the shift vector.  
The first long-lasting evolutions of distorted black holes
relied on a class of hyperbolic shift evolution schemes, known as
$\Gamma$-freezing conditions, which were inspired by
the BSSN formulation.
With this approach, a single distorted black hole could be evolved
indefinitely (e.g., for thousands of $M$) with reasonable accuracy 
\cite{Alcubierre:2001vm,Alcubierre:2002iq}. These studies
allowed the full numerical determination of gravitational
waveforms from black hole ringdowns as the distorted 
black hole settled down to a physically and 
numerically stable final state, and provided a foundation for future
advances in black-hole binary simulations \cite{Alcubierre:2002kk}.
 
\subsubsection{Problems with moving black holes}
\label{sssec:bhbmodel_evol_problems}

However, another aspect of 
spatial gauge conditions remained a critical issue for long-lasting 
black-hole binary simulations.  As general relativity allows arbitrary coordinate
systems, many groups adopted coordinate conditions which did not allow 
the black holes to move through the computational domain.  This simplified 
the problem of dealing with the black hole singularities, which were handled 
either by excising the black hole interiors (within the horizons)
 from the computational domain \cite{Thornburg87},  
or by representing the black holes as punctures \cite{Brandt:1997tf}.  
Though progress had been made toward implementing
moving excision regions \cite{Shoemaker:2003td}, it was computationally much 
simpler to demand that the excision region remain
fixed on the grid \cite{Alcubierre:2000yz}.   Similarly, the puncture 
treatment was formulated so that the black hole was fixed on the
grid and the singularity was
factored out and handled analytically
in a time-independent manner.  However, for binary configurations in which 
the black holes physically move, the cost of keeping the black holes
fixed in coordinate space was paid by the twisting and stretching of
the dynamical fields, eventually
leading to large computational errors. 

For inspiraling binary configurations one potential solution was to attempt
to untwist the geometric field data by imposing
comoving coordinates using a shift
vector that is dynamically adjusted during the evolution of the 
binary to minimize
motion of the black hole horizons \cite{Alcubierre:2004hr}.
With this approach, Br\"{u}gmann \etal achieved a significant milestone: 
the first  
simulation of a full orbital cycle of a black-hole binary system prior to
merger \cite{Bruegmann:2003aw}.  In this work, the authors applied a dynamically
adjusted corotating frame tracking the measured position of
the black hole apparent horizons.
This allowed a simulation which remained accurate for 
a little longer than the 
 $\gtrsim 100M$  duration of a complete orbit.
However, problems occurred with the large characteristic speeds that resulted
near the outer boundaries,  limiting the domains of these simulations and 
preventing gravitational waveforms from being measured effectively.  The 
coordinate control also required fine tuning which eventually failed, 
causing the simulation to crash as the black holes finally approached merger.

\subsection{Robust black-hole binary simulations}
\label{ssec:bhbmodel_robust}

\subsubsection{First Orbit and Merger simulation}
\label{sssec:bhbmodel_robust_first}

\begin{figure}
\includegraphics*[width=5.0in]{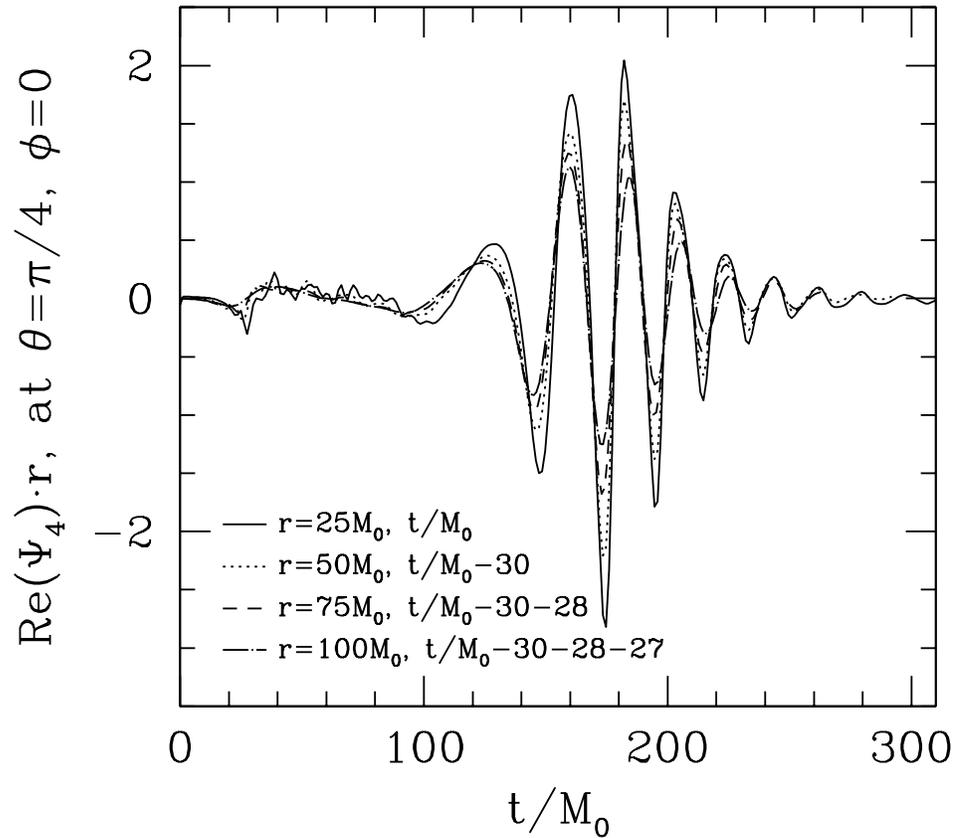}
\caption{The first gravitational waveform for binary of equal-mass
black holes evolving through
a single plunge orbit, merger, and ringdown, computed by Pretorius.
The waveforms were extracted at four radii from 
the source, and then
shifted in time to overlap for comparison.
Reprinted with permission
from \cite{Pretorius:2005gq} and copyright 2005 by the American Physical
Society
(http://link.aps.org/abstract/PRL/v95/e121101).}
\label{Fig2_PretoriusPsi4}
\end{figure}

In early 2005 Pretorius shocked the numerical relativity community by
announcing the first complete, robust simulations of an 
equal-mass 
black hole merger \cite{Pretorius:2005gq}.
After completing $\sim 1$ orbit, the black holes plunged and merged to 
form a single distorted black hole that then rings down.
Pretorius extracted the gravitational waves to obtain the
first inspiraling merger waveform directly from numerical relativity,
shown in Fig.~\ref{Fig2_PretoriusPsi4}.

Pretorius employed several techniques that were very
different from most other numerical approaches
to the black-hole binary problem \cite{Pretorius:2006tp}.  
 Rather than using the BSSN formulation, he applied a
generalized harmonic formalism \cite{Pretorius:2004jg}, directly
integrating the spacetime metric with evolution equations of
second-order in both space and time.  These 
equations were implemented numerically using 
adaptive mesh refinement to allow high resolution around the
black holes while maintaining a large computational domain.  Pretorius
utilized spatial coordinates compactified to draw spatial infinity
into the computational domain, with a choice of gauge-evolution
strongly tied to his evolution formalism.  Following \cite{Gundlach:2005eh},
he added terms to the evolution equations to specifically damp
away any violations of the constraints.  In his simulations, the black holes
were excised and moved freely across the computational grid. 
Pretorius' success with such novel methods
quickly raised questions as to whether the struggling, more widely-pursued
BSSN-based puncture approach might be off course.

\subsubsection{Moving Punctures}
\label{sssec:bhbmodel_robust_punctures}

\begin{figure}
\includegraphics*[width=5.0in]{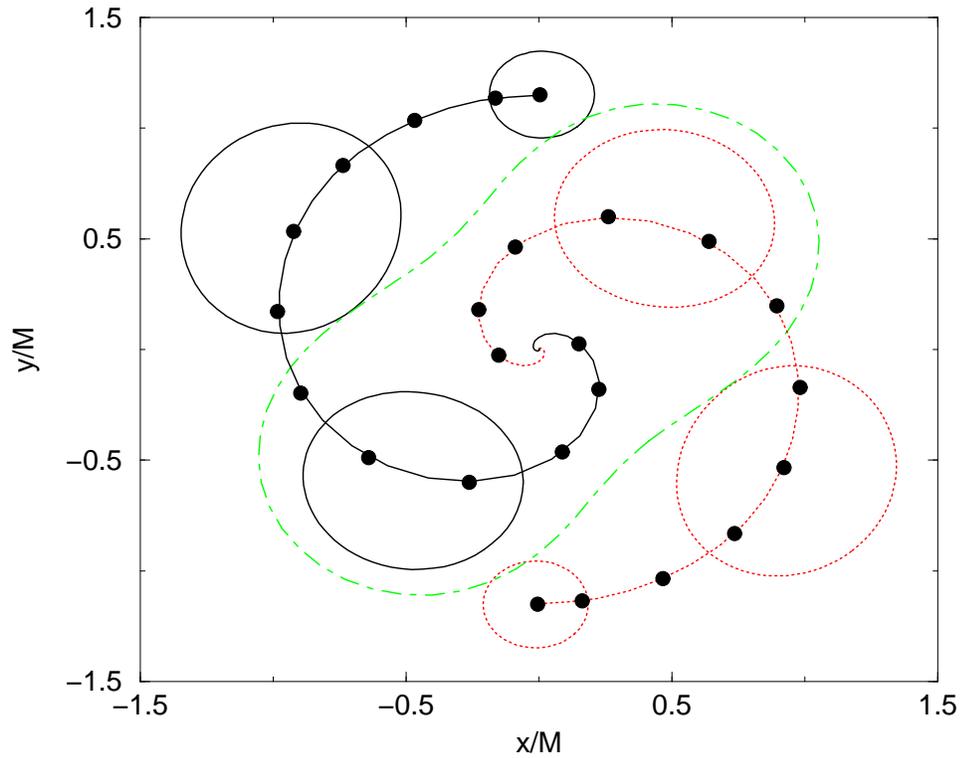}
\caption{Puncture trajectories from the merger of an equal-mass nonspinning
black-hole binary calculated by the UTB group. The 
apparent horizons of the two pre-merger holes, outlined in solid
(black) and red (dashed), expand due to coordinate effects. Also
shown is the first detected common horizon, outlined in green (dot-dashed);
this designates the point of merger and
has a ``peanut'' shape before it settles down.
Reprinted with permission
from \cite{Campanelli:2005dd} and copyright 2006 by the American Physical
Society
(http://link.aps.org/abstract/PRL/v96/e111101).}
\label{Fig3_UTBPRLtrack}
\end{figure}

Later that same year, however, a new robust method 
based on the BSSN formulation was announced.  The
``moving puncture'' method was discovered simultaneously and
independently by the groups at the University of Texas at Brownsville
 (UTB) \cite{Campanelli:2005dd} and NASA's Goddard Space Flight Center (GSFC)
\cite{Baker:2005vv}.
In this approach the black holes 
are represented as punctures, but are not constrained to remain
fixed on the coordinate grid. Rather,
they are allowed to move freely through the grid using novel
coordinate conditions.  
Figure~\ref{Fig3_UTBPRLtrack} shows the trajectories of the puncture
black holes as computed by the UTB group; the formation of a common
horizon marks the point of merger.
The moving puncture method
eliminates the analytic representation of the
puncture singularities in favor of an approximate numerical
treatment within the black hole horizons. 

 The UTB and GSFC groups had discovered and applied similar methods to 
the same problem: evolving an equal-mass nonspinning black-hole binary from
near the final orbit, 
through merger and ringdown, and studying the gravitational waves.
The first generation of merger waveforms from Pretorius, GSFC, and UTB showed
the same simple burst of radiation ending in a damped-sinusoidal ringdown,
and were qualitatively consistent 
with the Lazarus project results discussed above
and shown in Fig.~\ref{Fig1_LazarusWF}.

The discovery of the moving puncture method ignited the field 
of black-hole binary evolutions.  Since it was based on commonly used 
methods, most researchers in the field were quickly able 
to achieve accurate and stable evolutions using their existing codes,
with the adoption of 
simple coordinate conditions \cite{vanMeter:2006vi}.  
Suddenly, the game was on and
nearly all the groups were participating.

\section{The Physics of Black-Hole Binaries}
\label{sec:bhb}

With the advent of successful numerical evolutions of binaries that inspiral and
merge, the numerical relativity community's focus changed to
investigating the physics of binary mergers. This advancing frontier is 
enabled by continued technical improvements in areas ranging from 
initial data prescriptions to more accurate numerical methods. 

\subsection{Merger Dynamics and Waveforms}
\label{ssec:bhb_dynamics_waveforms}

The astrophysical black hole mergers that are the targets of current and
future gravitational wave detectors are expected to reach the merger stage
after 
having radiated away any initial eccentricity \cite{Peters:1963ux,Peters:1964zz}
and proceeding through
a long quasi-circular inspiral.
All equal-mass, nonspinning binary merger simulations starting
from such orbits in the inspiral should produce 
the same gravitational waveform, subject only to rescaling with 
the total mass of the system.
For many years, concerns had been raised about the accuracy and 
realism of black-hole binary initial data sets, 
including the effects of spurious 
gravitational radiation and eccentricity \cite{Lousto:1997ge,Damour:2000we,Pfeiffer:2002xz}.
With numerical relativity now able to simulate 
the final merger, the next step was to run models with enough 
orbits before the plunge and merger to get complete and 
reproducible waveforms starting from the late inspiral.

\subsubsection{Equal-Mass, Nonspinning Black Holes}
\label{sssec:equal-mass-nospin}

The GSFC group  
produced the first representation of the definitive 
waveform for the final stages of a merger of equal-mass, nonspinning black holes \cite{Baker:2006yw}.
They carried out a series of four simulations with the holes  starting 
from quasicircular
initial conditions at increasingly larger separations.
In these runs, the holes completed $\sim 1.8, 2.5, 3.6$, and $4.2$ orbits 
before the formation of a common horizon.
To compare the results of these models,
they chose the moment of peak gravitational radiation amplitude as 
the fiducial time $t = 0$.

The orbital dynamics of the binaries can be examined by tracking
the black hole centers, given by the location of the punctures.
The black hole trajectories
for each run were oriented so that they superpose at 
the fiducial $t=0$.
In the early stages of each run, the tracks clearly showed the effects of 
eccentricity in the initial conditions, with the amount of initial
eccentricity decreasing for more widely separated holes.
As the holes spiraled together deeper into the strong-field regime, the 
tracks locked on to a universal trajectory independent of their initial
conditions that continued for the 
last orbit, plunge, and merger.
 
\begin{figure}
\includegraphics*[width=5.0in]{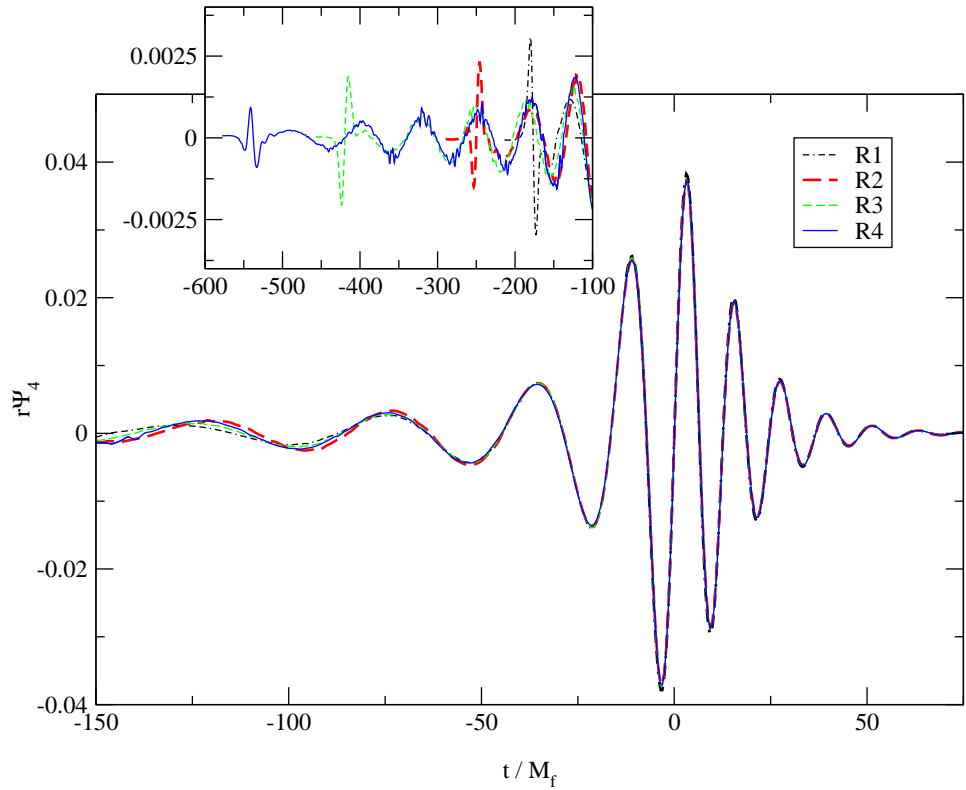}
\caption{The universal waveform for equal-mass, nonspinning
black holes calculated by the GSFC group. The figure shows waveforms from four
simulations with increasingly larger initial separations between the
black holes.  These waveforms were
shifted in time so that the peak radiation amplitude occurs at
$t = 0$. 
Reprinted
from \cite{Baker:2006yw} and copyright 2006 by the American Physical
Society
(http://link.aps.org/abstract/PRD/v73/e104002).}
\label{Fig4_QCNwaves}
\end{figure}

Figure~\ref{Fig4_QCNwaves} reveals the corresponding
universal gravitational waveform.  Here, 
the dominant $\ell=2$, $m=2$ quadrupolar component for each run
is shown, shifted in time so that the peak radiation amplitude occurs
 at $t = 0$.
Starting from $t = -50 M$ the waveforms show nearly 
perfect agreement, differing from each other at the level of $1\%$.
The signals for the preceding few orbits agree at the level 
of $\sim 10\%$, except for a brief burst of spurious radiation at 
the start of each run. 

Note that the merger waveform shows a remarkably simple shape, making a smooth
transition from the inspiral chirp to the damped sinusoid of the ringdown.
As the merger begins, both the wave amplitude and frequency increase,
albeit faster than in the inspiral.  The amplitude reaches a peak and
then decreases, dropping exponentially through
 the ringdown.  The frequency increases
monotonically to a maximum value that remains constant during the 
ringdown. 

Of course, the black-hole binary merger dynamics and waveforms can be
altered by the presence of large amounts of 
eccentricity \cite{Hinder:2007qu} and spurious
gravitational radiation \cite{Bode:2007dv} near the time of the plunge.
However, the robustness of the merger to modest deviations from astrophysical
initial conditions opened the door to studying many cases of interest
using relatively short simulations, starting just a few orbits before
the start of the plunge.

\subsubsection{Unequal-Mass, Nonspinning Black Holes}
\label{sssec:unequal-mass-nospin}

Astrophysical black-hole binaries are unlikely to have exactly equal masses.
Currently, numerical relativists
are able to evolve systems with mass ratios up to $q = 10$ 
\cite{Gonzalez:2008bi}.
Starting from quasicircular orbits the simulations 
show that the merger phase for nonspinning, unequal mass
black hole binaries is robust to modest
deviations from these initial conditions and produces a generally
simple waveform shape.

An important tool for analyzing these mergers is a decomposition
into spin-weighted spherical harmonic modes.
Berti \etal \cite{Berti:2007fi} analyzed a set of unequal-mass nonspinning
mergers with mass ratios ranging from $q = 1$ to $q = 4$.
Studying the multipolar distribution of the radiation, 
they found that the sub-dominant modes ($\ell > 2$)
become more important, carrying a larger fraction
of the energy, as $q$ increases.  Specifically, for $q > 2$, the
$\ell = 3$ mode typically carries $\sim 10\%$ of the radiated energy.
Also, as expected from symmetry considerations, the
odd-$m$ modes are
suppressed in the equal-mass limit. 

Baker \etal \cite{Baker:2008mj} carried out a complementary study of the
radiation from nonspinning mergers with mass ratios in the range
$1 \le q \le 6$.  The multipolar decomposition clearly shows that the
hallmark simplicity of the waveform persists for $q > 1$ and extends to 
each of the spherical harmonic components $\ell \ge 2$;
this property has recently
been shown to extend to the $q=10$ case \cite{Gonzalez:2008bi}.
In the full mode-summed waveform, this simple shape is also seen when 
viewing along the system's orbital axis, where the quadrupole mode
dominates.
  A somewhat more complex appearance arises
by viewing the system off-axis, where higher multipoles
contribute more strongly to the waveform at various angles. 

Throughout the entire coalescence, each of the spherical harmonic waveform
components is circularly polarized, 
with steadily varying phase and amplitude 
\cite{Baker:2008mj}.
For each mass ratio $q$, the rotational phase (and frequency) 
of the different 
$(\ell,m)$ components are the same. 
During the inspiral this is expected,
since the waveform phase is equal to the rotational phase multiplied
by the mode number $m$.
However, for the $\ell = m$ modes this relationship also holds throughout
the merger and into the ringdown.  

These properties suggest a simple conceptual interpretation in which 
the radiation is generated by an ``implicit rotating source.''
In this picture, each $(\ell,m)$ mode is generated separately by the
$(\ell,m)$ moment of some implicit source.  The fixed relationship
for the $\ell = m$ modes implies that these components of the source
rotate rigidly through the entire coalescence.  The $\ell \ne m$
components are less rigid and peel away from the main $\ell = m$
trend during the merger \cite{Baker:2008mj}.

\subsubsection{Mergers of Black Holes with Spin}
\label{sssec:unequal-mass-spin}

The mass ratio $q$ is a one-dimensional cut into the
parameter space of black-hole binaries. The remaining parameter space is dominated by
the spin angular momentum of each hole. As spin is a vector quantity,
this adds six more dimensions. 
Simulations of spinning black holes first focused on systems whose spins
were expected, on the basis of PN arguments, to have the
least effect on the orbital motion.
Binaries with aligned and anti-aligned spins are relatively easy to
treat as there is no precession of the orbital plane or the individual 
spins. When one or
both holes has a spin not parallel to the orbital axis, spin-orbit
and spin-spin interactions will cause precession that can
complicate the evolution and the
resulting waveforms \cite{Apostolatos:1994mx,Kidder:1995zr}. 

The first merger evolutions of spinning black holes were carried out
by Campanelli \etal \cite{Campanelli:2006uy}. They simulated the
mergers of two equal-mass highly spinning black holes
with $\ahat_{1,2} = .757$, and both spin vectors either aligned or
anti-aligned with the orbital angular momentum. 
Here, $a \le M$ is the magnitude of the black hole spin angular momentum
per unit mass.  
They also evolved a nonspinning equal-mass binary for comparison.
All three binaries had the same initial orbital angular frequency
 corresponding to an orbital period $\sim 125M$,
and merged to form a rotating remnant black hole with $\ahat_{\rm final} < 1$.
However, the
aligned system took noticeably more orbits to merge than the
others. This behavior is caused by the spin-orbit interaction,
which produces an effective  force between
the black holes, either an attractive or repulsive
for the anti-aligned or aligned cases, respectively.
All three binaries generate remarkably similar gravitational
waveforms having a simple shape, with the aligned case displaying
a longer wavetrain and the anti-aligned case a shorter one.

The first fully numerical evolutions of strongly precessing spinning
systems were carried out by
Campanelli \etal \cite{Campanelli:2006fy,Campanelli:2007ew}, who
observed both significant precession of the orbits and a ``spin flip,''
in which the final post-merger black hole spins in the direction opposite
to the two initial black holes.
More recent work is beginning to probe the effects of spin precession
on waveforms \cite{Campanelli:2008nk}.

Studies of this most general of (non-eccentric) parameter sets 
are still in the early stages. No systematic study of waveform
shapes and polarizations has been carried out yet.
A preliminary study of the multipolar structure of gravitational 
waves from several classes of binaries with equal spins
has been performed
by Berti \etal \cite{Berti:2007nw}.  They considered an equal-mass case with
aligned spins; several $q = 4$ binaries with antialigned (down-down)
spins; and three unequal-mass binaries with spins initially in the
orbital plane and pointing in opposite directions.  
Examining the distribution of gravitational-wave energy emitted by
various modes, they find that, as in the case of
nonspinning mergers, odd-$\ell$ multipoles are suppressed for $q = 1$
and that, as $q$ increases, more energy is radiated in higher-$\ell$
multipoles.

\subsection{The Spin of the Final Black Hole}
\label{ssec:bhb_final_spin}

Because the state of the final black hole formed from the
coalescence depends primarily on 
termination of the inspiral and the late burst of radiation in the merger,
it can be accurately probed with relatively short simulations of just
a few orbits.

The merger of two equal-mass nonspinning
black holes produces a final black hole with a moderately high
spin, $\ahat_{\rm final} \sim 0.69$
\cite{Baker:2002qf,Pretorius:2005gq,Campanelli:2005dd,Baker:2005vv,Scheel:2008rj}.
Since the black holes each start out with no spin, and any tidal spin-up
is negligible \cite{Campanelli:2006fg}, the spin of this final black hole arises
from the orbital angular momentum of the original binary.  
Simulations show that, for this simplest black hole merger, the final
spin is ``universal,'' or independent of the initial black hole
separation, for modest deviations from quasicircular initial conditions.

For mergers of nonspinning black holes with unequal masses, the spin of
the final black hole decreases as the mass ratio $q$ increases.  Simulations
with mass ratios up to $q = 10$ show that the final spin parameter scales 
as $\ahat_{\rm final} \sim q/(1+q)^2$, where $\ahat_{\rm final} \approx 0.48$
for $q = 4$ and  $\ahat_{\rm final} \approx 0.26$ for $q = 10$
\cite{Berti:2007fi,Baker:2008mj,Gonzalez:2008bi}.

The effects of spin-orbit and spin-spin coupling
can become important in determining the spin of the final black hole.  
In the simplest cases, the spin vectors are parallel to the orbital
angular momentum.  Depending on the mass ratio and the black hole spin,
the merger can result in a final black hole with a larger
(spun up) or a smaller (spun down) spin than either of the progenitors
\cite{Baker:2003ds,Campanelli:2006uy,Pollney:2007ss}.
In certain cases, the merger can lead to a spin flip with the final
black hole spinning in a direction opposite to the spins of the initial holes;
in particular, it is possible to produce a final nonspinning black hole,
$\ahat_{\rm final} = 0$, from the merger of two spinning holes
\cite{Buonanno:2007sv,Berti:2007nw}.  More
general black-hole binaries, with misaligned spins,
are further complicated with the effects of
precession \cite{Campanelli:2006fy,Tichy:2007hk,Dain:2008ck}.

Several attempts have been made to produce expressions for the final spin
vector using analytic techniques or by fitting to results from numerical
simulations \cite{Buonanno:2007sv,Kesden:2008ga,Rezzolla:2007xa,
Rezzolla:2007rd,Rezzolla:2007rz,Tichy:2008du,Lousto:2009mf}; see
\cite{Rezzolla:2008sd} for a review.
Of particular interest is the question of whether a black hole merger can
produce a maximally spinning hole ($\ahat_{\rm final} = 1$)
 or indeed exceed the Kerr limit ($\ahat_{\rm final} > 1$).  Current
research suggests that this is not possible, even for mergers
occurring from hyperbolic encounters \cite{Healy:2009ir} and mergers
of highly boosted black holes \cite{Sperhake:2009jz}.

\subsection{Recoil Kicks from Gravitational Radiation}
\label{ssec:bhb_kicks}

A notable phenomenon arising from asymmetric binary systems
is the merger recoil or kick -- a net movement of the
end-state black hole from the system's center of mass, caused by the
anisotropic emission of gravitational radiation during  the
coalescence. Fitchett \cite{Fitchett_1983} produced a useful formula for
the kick velocity, which has significant applications in
astrophysics, using a quasi-Newtonian approximation.  
Several authors also calculated the recoil analytically
using PN approximations
\cite{Favata:2004wz,Blanchet:2005rj,
Damour:2006tr,Wiseman:1992dv}. 
However, since the dominant part of the effect builds up
in the strong-field regime close to merger,
full numerical relativity simulations are needed for 
accurate calculations of the kick.

\subsubsection{Kicks from nonspinning black hole mergers}
\label{sssec:bhb_kicks_nospin}

In 2006, recoil from a fully numerical binary merger was demonstrated for
the first time by Herrmann \etal \cite{Herrmann:2006ks}, for plunging
black-hole binaries with mass ratios as large
as $q\sim 3.1$. This was followed soon
afterwards by a full orbit and plunge simulation by Baker \etal
\cite{Baker:2006vn}, who found a recoil of between 86 and 97 \kms
for a mass ratio $q=1.5$.

\begin{figure}
\rotatebox{-90}{\includegraphics[width=4.0in]{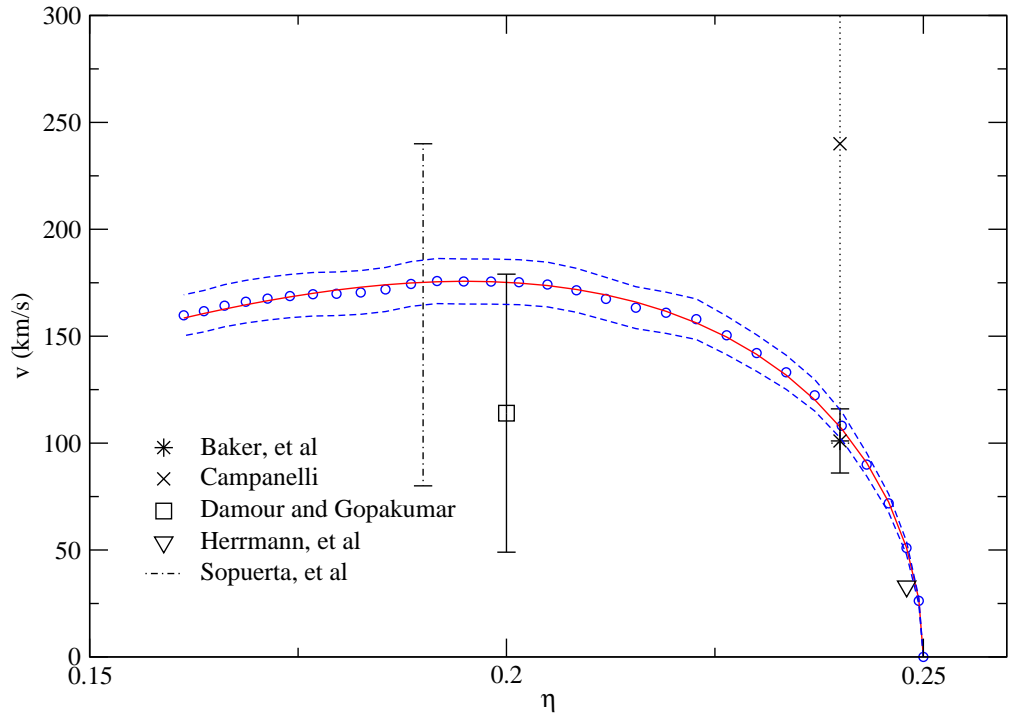}}
\caption{Recoil kicks for nonspinning black-hole mergers for mass ratios $q \in \{1.0,3.95\}$,
or $\eta \in \{0.161,0.25\}$; here, 
$\eta = q/(1+q)^2$.  The open circles are the results obtained 
from numerical relativity simulations by 
Gonzalez \etal\cite{Gonzalez:2006md},
with a functional fit given by the continuous (red) line, and uncertainties of $\sim 6\%$
indicated by the dashed (blue) lines above and below.
For comparison, earlier numerical
results (discrete symbols with error bars) are also shown.
Reprinted with permission
from \cite{Gonzalez:2006md} and copyright 2007 by the American Physical
Society
(http://link.aps.org/abstract/PRL/v98/e091101).
}
\label{Fig5_Jenakickfit}
\end{figure}

A more systematic study of recoil from mergers of unequal-mass 
binaries was produced by
Gonzalez \etal in 2007 \cite{Gonzalez:2006md}, who studied 
the merger of binary systems with mass ratios between 
$q=1$ and
$q=4$. 
Figure~\ref{Fig5_Jenakickfit} shows the resulting range
of recoil speeds, together with several earlier numerical and analytical
estimates. The authors synthesized these into a single recoil formula, a
nonlinear correction to the Fitchett formula, yielding a
maximum recoil of 175 \kms for a mass ratio of $q \sim 3$. 
This has
recently been tested for the more extreme $q=10$ mass ratio, with general
agreement \cite{Gonzalez:2008bi}.

\subsubsection{Spinning black hole mergers and superkicks}
\label{sssec:bhb_kicks_spin}

Calculations of recoil in the much larger parameter space of spinning binaries
began with the non-precessing cases of holes with spins aligned (or
anti-aligned) with the orbital angular momentum. Several studies of this region
of parameter space \cite{Herrmann:2007ac,Koppitz:2007ev,Pollney:2007ss}
have revealed that the kick velocity has 
a quadratic dependence on initial spins, with a maximum kick
of 448 \kms for extremal Kerr holes,
$(a/M)_{1,2} = 1$. For these anti/aligned cases, as well as
for nonspinning unequal-mass black-hole mergers, the 
direction of the kick velocity is
always in the orbital plane.

Meanwhile attention turned to more general black hole
spins.  Campanelli \etal \cite{Campanelli:2007ew}
speculated from PN
arguments that optimal spin configurations could 
give rise to huge ``superkicks,'' with velocities 
$> 1000$ \kms, out of the initial orbital plane. The first such superkick ---
around 2500 \kms --- was soon observed by Gonzalez \etal \cite{Gonzalez:2007hi}.
Tichy \etal \cite{Tichy:2007hk}
have argued that superkicks arise in mergers with general spin
orientations,
while greater insight into the mechanism
of these kicks has been developed \cite{Brugmann:2007zj,Schnittman:2007ij}.

With such a large parameter space to cover, it seems useful to try to construct a
general formula that will describe kicks from arbitrarily spinning binaries.
Baker \etal \cite{Baker:2007gi} used new and existing aligned-spin results to produce
a single unifying formula for in-plane kicks; Campanelli \etal \cite{Campanelli:2007ew,Campanelli:2007cg}
proposed an extension to this model, with scaling of the superkick out-of-plane
contribution motivated by PN theory. The leading dependence in this formula
on the angles between spins and linear momenta of the pre-merger holes has strong support
from numerical simulations
\cite{Campanelli:2007cg,Brugmann:2007zj,Lousto:2007db}; however, the dominant scaling with mass
ratio is still in dispute \cite{Baker:2008md,Lousto:2008dn}.

\subsection{Longer Waveforms: Modeling the Late Inspiral}
\label{ssec:bhb_longwfs}

Many key features of black-hole binary interactions can be modeled usefully with only a small handful of binary
orbits before merger. Quantities such as radiated energy and momentum are bulked near the merger
and have been shown to be robust to the addition of one or two extra orbits (see Section \ref{sssec:equal-mass-nospin}).
However, optimal observational 
studies of gravitational waveforms ultimately require theoretical predictions
for the full waveform, starting at large separations during the inspiral.
  
Before the breakthroughs in numerical relativity, most information about the dynamics of compact binaries
came from PN theory. 
This approach supplied the particle trajectories, energy flux, and -- most
importantly for detector scientists -- the gravitational waveforms themselves. However PN theory fails before the
system merges, so the waveforms are necessarily incomplete. With the advent of  
numerical simulations encompassing many orbits,
scientists finally have a way to develop complete information about the waveforms.
This requires longer simulations which reveal the last part of the binary inspiral, and
allow overlap with PN waveform predictions.

\subsubsection{Low-eccentricity initial data}
\label{sssec:bhb_longwfs_lowecc}

Simulations for these studies require careful attention to the initial 
configuration of the black holes.
For a significant population of astrophysical binaries, it can be expected 
that the orbits have circularized through gravitational-wave emission
over many orbits prior to merger.  Numerical simulations try to mirror this expectation by selecting
initial momenta consistent with near-zero eccentricity for the relatively
small separations at which a full numerical relativity simulation becomes feasible.

For equal-mass, nonspinning binaries, two basic approaches have proved
effective in reducing spurious eccentricity. The more direct
method is to model the observed eccentricity, and then
adjust the momenta using a Newton-like step to zero it out; this
approach has led to extremely low eccentricities \cite{Boyle:2007ft,Scheel:2008rj}. An alternative approach,
adopted by Husa \etal \cite{Husa:2007rh}, is to use the PN Hamiltonian equations of motion to model the early
evolution of the particle trajectories, starting from large ($\sim 50 - 100M$) separations; the emission of
radiation during this process naturally circularizes the orbit, and low-eccentricity momenta can be read off
at the desired separation. This approach has recently been tested with spinning binaries as
well \cite{Campanelli:2008nk}.

The first long waveform was produced by the
 GSFC group \cite{Baker:2006ha,Baker:2006kr}
for equal-mass,
nonspinning black holes starting $\sim 7$ orbits or $\sim 14$
gravitational wave cycles before merger. 
Baker \etal
estimated their numerical errors in waveform phase from \cite{Baker:2006kr} as a function of frequency, finding
that above a certain frequency numerical errors are smaller than internal
errors in the PN sequence. 
In addition, the numerical and PN waveform phases 
agree to within one radian of phase drift 
for a little over ten gravitational wave cycles preceding the last orbit 
before merger, comparable to numerical error estimates.
Hannam \etal \cite{Hannam:2007ik} improved on this
work, with phase and amplitude comparison between their low-eccentricity higher-resolution
evolutions and PN waveforms.

\subsubsection{Evolutions with Spectral Techniques}
\label{sssec:bhb_longwfs_spectral}

Numerical simulation codes based on pseudospectral differencing techniques
are particularly well-suited to long waveform studies.  
The Caltech-Cornell group adopted the constraint-damped generalized harmonic 
formalism used by Pretorius, an important component in developing a stable 
spectral evolution code \cite{Lindblom:2005qh,Scheel:2006gg}.
Also like Pretorius, their code handles black holes by 
excising the black hole interiors.  Their spectral code also employs 
a numerical grid that 
tracks the black holes explicitly.  Though there can be difficulties with
the changing geometry as black holes merge, this approach allows efficient
study of the long-lasting inspiral waveforms.

\begin{figure}
\includegraphics*[width=6.0in]{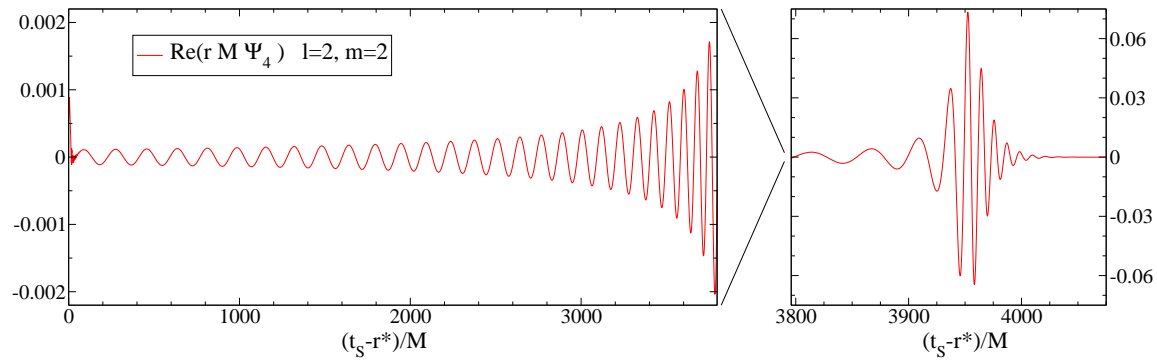}
\caption{The longest gravitational waveform for an equal-mass, nonspinning
black-hole binary merger computed by the Caltech-Cornell group.  
The left panel shows the early stages of the
waveform, during the inspiral.  The right panel displays the merger and
ringdown portions of the waveform.
Reprinted with permission
from \cite{Scheel:2008rj} and copyright 2009 by the American Physical
Society
(http://link.aps.org/abstract/PRD/v79/e024003).
}
\label{Fig6_CCWF}
\end{figure}

Recently this approach has provided, several of the longest and most accurate 
black-hole binary evolutions.\cite{Scheel:2008rj,Chu:2009md}
In particular, the Caltech-Cornell group have used their spectral code 
to simulate an equal-mass nonspinning binary
starting 16 orbits and 32 gravitational wave cycles before merger: 
see Fig.~\ref{Fig6_CCWF} \cite{Scheel:2008rj}
These much longer waveforms have been used to validate several competing
PN models \cite{Boyle:2007ft}. 

\subsubsection{Comparing results}
\label{sssec:bhb_longwfs_compare}

With so many pre-merger waveform cycles now available for the 
equal-mass case, the results can be cross-checked by comparing 
the ``complete'' waveform -- inspiral, merger and ringdown -- 
between groups. This is important to verify expectations that
differences in methodology and residual numerical effects, 
such as unwanted eccentricity, are unimportant.  A recent effort,
dubbed the ``Samurai project,'' analyzed
long waveforms from several groups in the light of detectability 
criteria for the LIGO and Virgo ground-based gravitational-wave detectors.
They found that the available numerical relativity
waveforms are indistinguishable in these detectors for signal-to-noise ratios (SNRs)
$\lesssim 25$ \cite{Hannam:2009hh}.

\subsubsection{Long Waveforms For More General Black Holes}
\label{sssec:bhb_longwfs_generic}

Hannam \etal have investigated the properties
of highly spinning orbit-aligned black holes, comparing their 
phase and amplitude with PN predictions over
the last ten waveform cycles \cite{Hannam:2007wf}. In the best 
cases, they find  phase agreement within 2.5 rad
for 3.5PN, and amplitude agreement to within around 12\% with 
restricted
PN. While not at the same level as their
nonspinning results \cite{Hannam:2007ik}, this may be attributable 
to the lower orders of accuracy available
for spinning black holes in PN theory at the present.  The 
Caltech-Cornell group has recently conducted 
long-lasting simulations of black holes with spins aligned and 
anti-aligned with the orbital angular 
momentum, calibrating a tunable PN waveform model to match the 
results \cite{Chu:2009md,Pan:2009wj};
see Sec.~\ref{ssec:outlook_analyticmodels}.  

Generic mergers involving non-aligned, and thus precessing, spins adds four new degrees of freedom to the problem.
A systematic understanding of waveforms generated by generic mergers will require considerably more study.  
Recent simulations
have begun to explore generic examples \cite{Campanelli:2008nk}.

\section{Applications in Astrophysics}
\label{sec:astrophys}

We have described a sampling of the explosion of numerical relativity
studies revealing some of the details of black-hole binary physics
as implied by General Relativity.  While more remains to be learned,
this new understanding is already making important contributions in 
planning and interpreting astrophysical black hole observations where 
Einstein's theory is applied as the standard model of gravitational physics.  

\subsection{Waveforms for Gravitational-Wave Observations}
\label{ssec:astrophys_da}

Experimental gravitational-wave detectors were first developed more than 40 years ago.
Although advances in design have increased their
sensitivity by many orders of magnitude, the extreme weakness of expected astrophysical
signals (strain amplitudes $\delta L/L \sim 10^{-21}$) means the observational challenge
is still huge.

The output of any gravitational wave
 detector will be a data stream that must be combed through to find
real signals. This search requires accurate ``template'' waveforms representing our
best picture of the radiation from expected sources; these templates can then be compared
with the data stream through matched filtering. We refer the reader to a review on
gravitational-wave astronomy for an overview of these techniques \cite{Camp:2004gg}.

Crucially, the most important sources of gravitational radiation
are expected to include the mergers of
black-hole binaries. Before the advent
of numerical relativity simulations of black hole mergers, the only test
waveforms available for use in data analysis studies 
were based on PN theory.  These waveforms were essentially 
only inspiral chirps and did not include the strong-field merger.
 The new, richer, information now available 
from numerical relativity has
revolutionized the data analysis picture in several ways.

\subsubsection{Detecting black-hole mergers}
\label{sssec:astrophys_da_detect}

\begin{figure}
\includegraphics[width=5.0in]{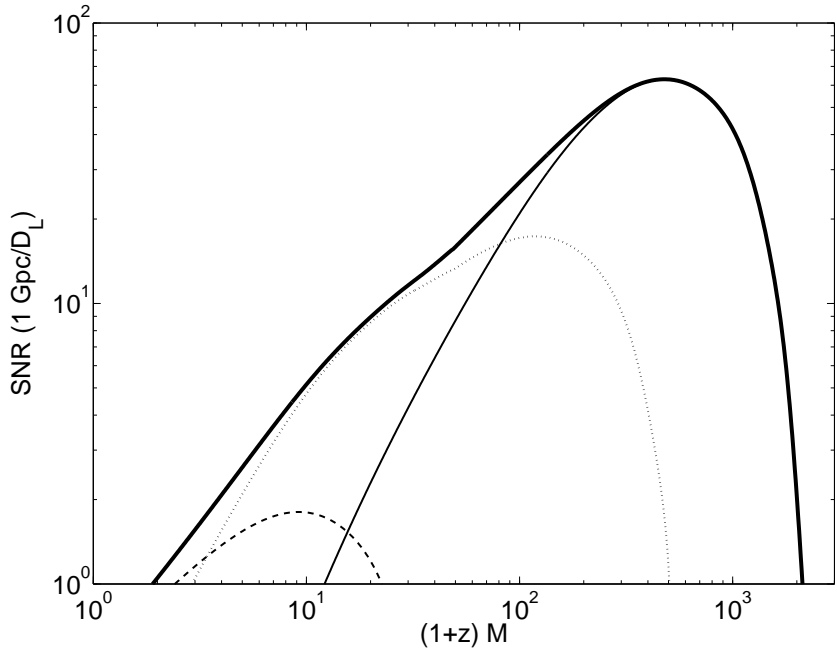}
\caption{Gain in SNR by including the merger segment of the waveform
for an equal-mass, nonspinning binary. 
The SNR for sources located at luminosity distance $D_L = 1$ Gpc is
plotted vs. (redshifted) mass for the Advanced LIGO detector. 
The dashed line shows the SNR calculated using PN techniques for the 
early inspiral part of the waveform, $ -\infty < t < -1000M$.
The dotted line shows the SNR using the late inspiral,
$-1000M < t < -50M$, which is the transition region from PN to
numerical relativity.  The SNR for the strong-field merger and
ringdown, $-50M < t < \infty$, was calculated using waveforms from a
numerical relativity simulation and is shown using a thin solid line.
Finally, the SNR from the entire waveform is given as the thick solid line.x
Here, $t=0$ marks the time of maximum gravitational radiation amplitude.
Reprinted
from \cite{Baker:2006kr} and copyright 2007 by the American Physical
Society
(http://link.aps.org/abstract/PRD/v75/e124024).
}
\label{Fig7_MergerObsALIGOSNR}
\end{figure}

\begin{figure}
\includegraphics[width=5.0in]{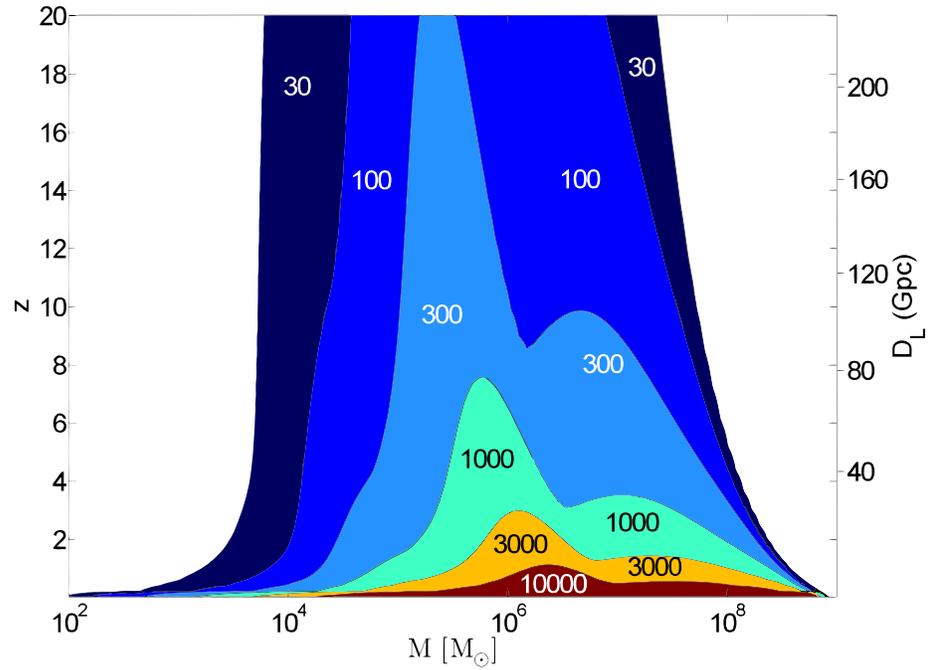}
\caption{Contours of SNR as a function of redshift $z$ and total
binary mass $M$ are shown for the LISA detector.  These have been calculated using
a hybrid equal-mass nonspinning waveform: using PN for the early inspiral, 
matching to a numerical relativity waveform in the late inspiral,
and continuing with the numerical waveform
through the merger and ringdown.
Reprinted
from \cite{Baker:2006kr} and copyright 2007 by the American Physical
Society
(http://link.aps.org/abstract/PRD/v75/e124024).}
\label{Fig8_MergerObsLISASNR}
\end{figure}

The availability of the plunge-merger-ringdown portion of the
signal can greatly increase the SNR in the detector.
Armed with a long numerical merger waveform of acceptable accuracy, we can extend it backwards
to cover an arbitrarily long inspiral by matching to a PN waveform. 
Such a ``hybrid'' waveform was first produced
by the GSFC group for the equal-mass, nonspinning case \cite{Baker:2006kr}.
Using this hybrid, we can investigate the total achievable SNR, and the related
distance reach, for current and future detectors. 
Figure~\ref{Fig7_MergerObsALIGOSNR} demonstrates the gain in SNR from including
the merger portion of the waveform
for the ground-based Advanced LIGO detector.
Contours of SNR as function of redshift $z$ and total binary mass $M$
for the LISA detector are shown in 
Figure~\ref{Fig8_MergerObsLISASNR}.

Full numerical waveforms, and the longer hybrid waveforms generated from them, can
also be used to improve  existing data-analysis techniques and template sets. 
Since previously developed gravitational wave data analysis algorithms were not
based on knowledge of the merger waveforms, an obvious first step is to test how successfully
these techniques detect the merger signals predicted by numerical simulations. 
In 2009, the NINJA project \cite{Aylott:2009ya} used direct injection of a broad range of
short and long numerical merger waveforms into mock LIGO and Virgo data streams
for this purpose. The result was the most realistic testing ground to date for 
disparate data-analysis methods, including full and partial waveform template matching,
as well as unmodeled burst searches. Further studies of detection algorithms with numerical
relativity waveforms are continuing. 

\subsubsection{Measuring black-hole binary parameters}
\label{sssec:astrophys_da_paramest}

The SNR is only a crude guide to the specific detector response to gravitational
waveforms, however. The merger portion of the waveform, though short in duration,
may contain important new information not present in the inspiral. 
As a simple example,
we expect the time of merger itself to be well localized with the full waveform,
whereas it is not well-defined in the inspiral-only signal.

More generally, the different $(\ell,m)$ modes of a binary's full 
inspiral-merger-ringdown waveform scale differently with 
the time to merger. Modes that were not
significant during inspiral suddenly become more prominent
in the merger, and the detailed information
they carry becomes available to the observer \cite{Berti:2007fi,Baker:2008mj}

After detection of gravitational waves from distant sources, we are 
most interested in
identifying the physical parameters of the sources. Each of the seven intrinsic parameters
of a black-hole binary (mass ratio and spin vectors) will, in general, be imprinted
on the gravitational-wave signal, 
along with some extrinsic parameters such as sky position
and distance to the source. For sufficiently strong SNRs, expected for massive binary
mergers seen by LISA, we can expect to be able to extract some of these parameters at high precision.
While these 
parameters can be partially disentangled using inspiral-only 
template information, it has recently been found that the full merger waveform
can help break parameter degeneracies and hence drive down uncertainties in several
important physical parameters. 
Good localization of the source
on the sky
is especially important
for the development of multi-messenger astronomy. 
The recent parameter estimation studies of non-spinning mergers
that include the merger waveforms
indicate that LISA will be able to locate sky positions
within a few arcminutes for binaries with $\sim 10^6\MSun$ at cosmological
distances (redshift $z=1$)\cite{McWilliams:2009bg}.

\subsection{Consequences of Merger Recoil}
\label{ssec:astrophys_recoils}

As previously discussed, asymmetries in a black-hole binary system due to unequal masses and/or
spins result in the anisotropic emission of 
gravitational radiation, ultimately imparting a recoil to the merged remnant black hole.
Numerically it has been found that, for certain configurations of black hole spins, the recoil velocity can 
exceed the escape velocity of many galaxies.  
Since it is important
to determine the theoretical probability that a
massive black hole might be ejected from its host galaxy,
there have been some preliminary calculations based on simple distributions of
spins and mass ratios, and analytic fits of numerical results giving the dependence of the recoil on mass and spin
\cite{Schnittman:2007sn,Baker:2008md}.  Although a consensus on the exact probability of galactic ejection has yet to be achieved, 
there is general agreement that it is non-negligible.

Such rogue black holes may have already been observed in the form of two rapidly moving quasars
\cite{Shields:2009jf,Komossa:2008qd}.  It has been speculated that these particular quasars originated from the mergers of 
massive black holes during the coalescence of their host galaxies.  However a previous study of quasar data
found no indications of such recoil events, suggesting that they are rare \cite{Bonning:2007vt}.

The effect of recoil may also be observed less directly in its effect on the growth rates of black holes.  Those
ejected into the sparse intergalactic medium are less likely to encounter and merge with other black holes
\cite{Sesana:2007sh,Volonteri:2007et}.  Even the growth of recoiled black holes remaining within their host galaxies may be
affected, as the motion of the black hole can modify the rate at which it 
accretes matter \cite{Blecha:2008mg}.

\subsection{Mass and Spin Evolution}
\label{ssec:astrophys_massspin}  

The expected distribution of
masses and spins of astrophysical black holes is another topic of
considerable interest in astrophysics. 
In our current understanding, black holes grow from smaller ``seeds'' early in
the history of the universe through a combination of mergers and
the accretion of gas \cite{Sesana:2004sp}.  In general, most of the mass growth
is believed to come from accretion, with mergers providing a modest increase.

However, the gravitational radiation emitted during black hole coalescence
carries away energy, reducing the overall system mass by roughly several
percent.
The bulk of this loss happens quickly, in the final plunge and
merger stage of the coalescence.  With this rapid mass loss, nearby matter 
in an accretion disk around the remnant black hole might react
to the accompanying
 sudden change in gravitational potential and produce
a visible change in its electromagnetic
profile -- a possible electromagnetic counterpart to the burst of gravitational radiation  
\cite{2007APS..APR.S1010B,O'Neill:2008dg}.

For the final spin the situation is considerably more interesting, 
and may answer questions about double-jet ``X-shaped'' radio sources
\cite{Merritt:2002hc,Barausse:2009uz}. 
We have seen in Sec.~\ref{ssec:bhb_final_spin} that the final
spin from black holes merging in vacuum depends on the magnitudes of the initial spins and their
orientations relative to the orbital axis. 
The RIT group \cite{Lousto:2009mf} has recently used their extended mass and spin
formulas in a spin-evolution study, obtaining asymptotic spin distributions for
BH merger remnants assuming no accretion.
However, interaction of the merging holes with
surrounding gas may serve to align the binary spins before merger, changing the picture
somewhat \cite{Bogdanovic:2007hp}. In general, the effects of binary merger and accretion
have to be studied together for a coherent picture to emerge \cite{Berti:2008af}.

\section{Outlook}
\label{sec:outlook}

\subsection{Complete Analytic Waveform Models}
\label{ssec:outlook_analyticmodels}

Gravitational wave observatories may be sensitive to
hundreds or thousands of wave cycles.
Analysis of the observed data requires comparisons with
model signals representing the full variety of possible mergers.
Computing so many orbits of pre-merger evolution
using numerical relativity would be 
computationally very expensive.

The PN approximation
provides accurate representations of the dynamics and waveforms for the
long-lasting inspiral portion of the coalescence during which the black holes
remain fairly well-separated and their velocities remain relatively low.
The most valuable waveform models for gravitational wave data analysis
must combine the efficiency of the PN approach, while accurately representing
the final merger portion of the which is only understood by numerical 
simulations. This requires a means of analytically encoding the merger signals.

\begin{figure}
\includegraphics[width=0.85\textwidth]{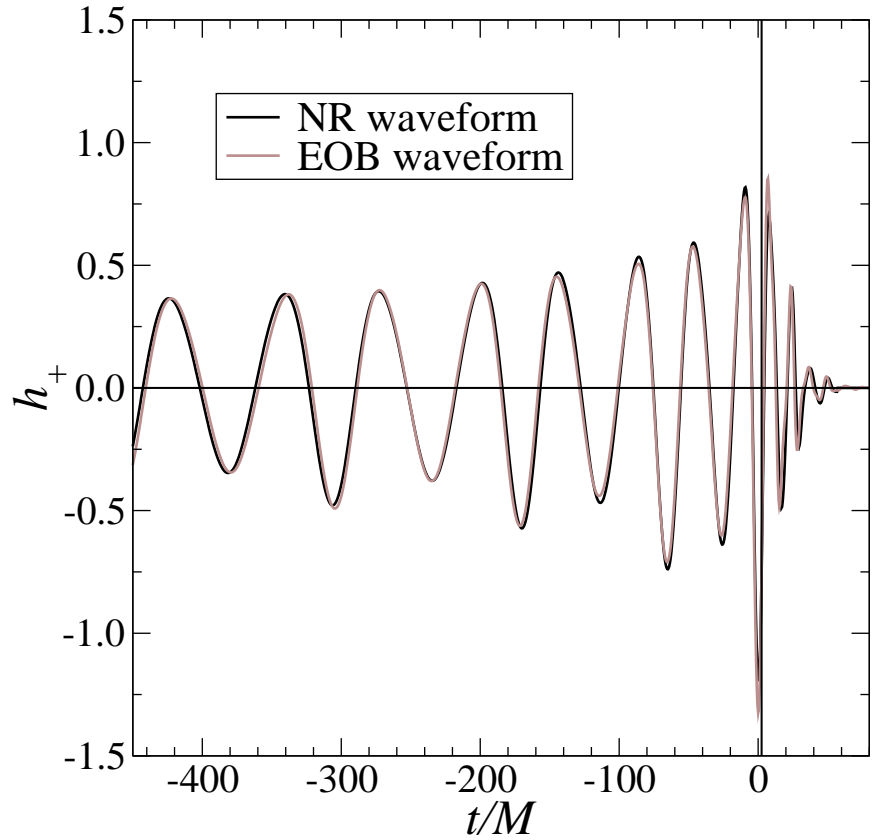}
\caption{Comparison of merger waveforms for a $q=4$ mass ratio 
nonspinning black hole binary calculated using the
analytic ``effective one-body'' (EOB) model and using numerical relativity
(NR).
Reprinted with permission
from \cite{Buonanno:2007pf} and copyright 2007 by the American Physical
Society
(http://link.aps.org/abstract/PRD/v76/e104049).
}
\label{Fig9_BuonannoEOBWF} 
\end{figure}

Several approaches are currently being explored for constructing these complete
signal models. Some are based on the 
analytic ``effective-one-body'' model of binary coalescence
\cite{Pan:2007nw,Buonanno:2007pf,Damour:2008te}.
Figure~\ref{Fig9_BuonannoEOBWF} shows a comparison between an effective-one-body
waveform for the merger of a nonspinning black hole binary with
mass ratio $q = 4$ with a numerically
simulated waveform \cite{Buonanno:2007pf}.
For specific cases, such models have been tuned to high 
accuracy to agree to agree with numerical
results \cite{Damour:2009kr,Buonanno:2009qa}.
Another approach, which models the phenomenological shape of hybrid waveforms
in frequency space \cite{Ajith:2007kx}, has been extended for the dominant
waveform modes from spinning, but non-precessing, mergers \cite{Ajith:2009bn}.
Further development of these models, and the production of a family of
simulated waveform spanning the parameter space, is a current focus 
of broad-based research collaborations.

\subsection{Improved Numerical Methods}
\label{ssec:outlook_methods}

\subsubsection{Initial Data}
\label{ssec:outlook_methods_id}

The initial data models currently
 used to begin numerical simulations do not 
perfectly represent the intended astrophysical systems.  Research continues to
improve these models.
For example, although some astrophysical black
holes are expected to have near-extremal spins, common initial data cannot represent holes 
with $\ahat_i \gtrsim 0.93$ \cite{Lovelace:2008tw,Dain:2002ee};
novel methods are being developed to go beyond this limit \cite{Lovelace:2008tw} .
Also, current initial data models generally do not contain
the physically appropriate radiation content 
for an inspiraling binary \cite{Damour:2000we,Pfeiffer:2005zm}, 
but rather harbor spurious
radiation that is not astrophysical\cite{Hannam:2006zt}.
Efforts are underway to include initial radiation content
more consistent with PN predictions \cite{Kelly:2007uc}.  
In addition, in the case of moving punctures,
the initial coordinates typically undergo a rapid transition 
at the start of the evolution
as the black holes relax into a more stable solution.
While the resulting transient pulse of ``gauge 
radiation'' does not alter the physics, it does contain fairly
high frequencies that can be challenging to resolve and it
has motivated construction of analytic initial coordinates that are closer to the numerically
evolved coordinates \cite{Hannam:2009ib}. 

\subsubsection{Evolution}
\label{ssec:outlook_methods_evol}

More efficient and accurate methods of numerical evolution would make simulations of
many binary orbits, particularly with large mass ratios, more computationally practical.
For representing spatial derivatives on a computational grid, 
the highest accuracy of finite differencing
stencil yet achieved in the context of numerical relativity is 8th order
in the grid spacing (although the accuracy of
interpolation between refinement boundaries is not yet commensurate)
\cite{Lousto:2007rj,Pollney:2009yz}.
Spectral methods are generally more accurate but less robust
than finite differencing,
requiring fine tuning for generic black hole binaries \cite{Szilagyi:2009qz}; they are
also unlikely to handle shocks in accreting matter (see
Section~\ref{ssec:outlook_matter}).  Alternatives to both finite differencing and
spectral methods, such as finite element methods, are also being investigated
\cite{Zumbusch:2009fe}.  Meanwhile more efficient time-integration techniques allowing
larger step sizes are being explored \cite{Lau:2008fb}.

\subsubsection{Wave Extraction}
\label{ssec:outlook_methods_extract}

Because the physical domain of a typical simulation is finite, gravitational radiation is usually extracted on a
sphere of finite radius rather than at infinity.  If multiple extraction surfaces are employed in a region of sufficiently high spatial 
resolution, then the
radiation can be extrapolated to spatial infinity.  A more accurate method
known as ``Cauchy-characteristic extrapolation'' extracts the gravitational
wave data at a finite radius and then evolves it along null geodesics to
future null infinity;  this method is
currently under development \cite{Reisswig:2009rx,Babiuc:2008qy}.

\subsection{Including Gas and Magnetic Fields}
\label{ssec:outlook_matter}

In addition to being gravitationally ``loud", black hole mergers may
also be electromagnetically visible.  Massive black holes at the centers of
galaxies are typically surrounded by gaseous accretion disks
and magnetic fields.  When the black holes merge,
the dynamics of the gas and magnetic fields may produce 
electromagnetic signals, counterparts to the emitted
gravitational radiation.  
For example, the inspiral may ``twist" the fields,
resulting in characteristic electromagnetic 
radiation \cite{Palenzuela:2009yr,Palenzuela:2009hx}
as well as heating of a surrounding accretion disk \cite{Reynolds:2006uq}.  
In addition, the violent merger dynamics may induce 
shock waves in accreting matter, in turn generating electromagnetic radiation.
The recoil of the merged remnant, in particular, may have such an effect on 
the accretion disk 
\cite{Armitage:2002uu,Milosavljevic:2004cg,Dotti:2006zn,Kocsis:2005vv,
Phinney:2007,2007APS..APR.S1010B,Kocsis:2007yu,Shields:2008va,Lippai:2008fx,
Schnittman:2008ez,Kocsis:2008va,
Haiman:2008zy,O'Neill:2008dg,Haiman:2009te,Chang:2009rx,
Megevand:2009yx}. 

\subsubsection{Multimessenger Astronomy}
\label{sssec:outlook_matter_motive}

Detection of electromagnetic counterparts of gravitational waves would be of great scientific value.
Current models of the complex merger physics (e.g. \cite{Balbus:1991ay}) could be directly tested.  Einstein-Maxwell theory,
the coupling of gravitational and electromagnetic fields on macroscopic scales, could be
verified.  In particular, equality of the speed of gravity with the speed of light could be confirmed \cite{Kocsis:2007yu,Palenzuela:2009yr}.
Astronomy would also benefit enormously, as the location and characteristics of gravitational wave sources could 
be corroborated.
In addition, electromagnetically visible mergers could serve as ``standard candles", beacons by which to measure
the accelerating expansion of the universe, while simultaneously playing the role of ``standard sirens" emitting gravitational
radiation.  Cross-correlating these signals could result in measurement of the cosmological ``dark energy" to unprecedented accuracy 
\cite{Lang:2008gh,Kocsis:2007hq,Jonsson:2006vc,Dalal:2006qt,Kocsis:2005vv,
Holz:2005df,Kocsis:2007yu,Arun:2008xf}.

\subsubsection{modeling matter}
\label{sssec:outlook_matter_model}

There has been some preliminary work on the dynamic effects of
the spacetime of a coalescing binary on surrounding matter.
 Modeling the accretion disk as geodesic particles, large collision
energies were found as the binary merged \cite{vanMeter:2009gu}.
And hydrodynamically simulating a gas cloud around around the binary,
luminosity due to shocks has been calculated \cite{Bode:2009mt}.
The generation of electromagnetic radiation by more direct means has
also been simulated, via the twisting of a magnetic field anchored in
the accretion disk, as it is frame-dragged by the binary
\cite{Mosta:2009rr,Palenzuela:2009hx,Palenzuela:2009yr}.
Future efforts will employ magnetohydrodynamic methods, 
where challenges include adequately resolving shocks and turbulence
in a dynamic spacetime, and accurately
representing the divergence-free magnetic field on an adaptively refined grid.

\section{Conclusion}
We hope to have conveyed some of the excitement of recent
progress in understanding black-hole binary physics, and
the applications of this knowledge in astrophysics.
These advances are the result 
of sustained efforts by a broad scientific community over many
years. In a brief review, it is impossible to adequately 
represent all of the excellent work that has contributed to 
the current state of knowledge.  We have only been able to 
provide a few of the highlights as seen through the lens of 
our particular perspective. 

We encourage interested readers to
pursue the subject further.  Other resources are available on topics
including: numerical relativity techniques 
\cite{Bona05,Alcubierre08},
the breakthroughs in black hole merger simulations \cite{Pretorius:2007nq},
black hole simulations for gravitational wave data analysis \cite{Hannam:2009rd},
and gravitational wave science generally \cite{Sathyaprakash:2009LR,Camp:2004gg}.

\section*{Acknowledgments}
We acknowledge support from NASA grants
06-BEFS06-19 and 08-ATFP08-0126. BJK was supported in part
by an appointment to the NASA Postdoctoral Program at
the Goddard Space Flight Center, administered by Oak
Ridge Associated Universities through a contract with
NASA.

%\bibliographystyle{../../bibtex/arnuke_revised}
%\bibliography{../../bibtex/references}

%% FIGURES 

%

\end{document}